\documentclass[12pt]{iopart}
\usepackage{iopams}
 \begin{document}

\title{Dark matter in dwarf galaxies}

\author{Matts Roos}

\address{Department of Physics\\ FI-00014 University of Helsinki, Finland}
\ead{matts.roos@helsinki.fi}

\begin{abstract}
Although the cusp-core controversy for dwarf galaxies is seen as a problem, I argue that
the cored central profiles can be explained by flattened cusps because they suffer from conflicting measurements and poor statistics and because there is a large number of conventional processes that could have flattened them since their creation, none of which requires new physics. Other problems, such as "too big to fail", are not discussed.
\end{abstract}

\maketitle

\section{Introduction}

The relativistic phenomenon of light rays bending around gravitating masses has
become an important tool to study massive objects in the Universe, such as galaxies,
supergalaxies and in general the distribution of mass on cosmological scales. A
surprising discovery has been that the Universe is filled with some unknown kind
of non-radiant dark matter (DM), which actually is transparent rather than dark. All
galaxies are embedded in halos of this non-radiant matter which extends far outside
the luminous baryonic matter, and their very kinematics is influenced by their DM
halos. The remarkable fact is that the DM fraction of the total energy content in the
Universe, $\Omega_c = 0.264$ \cite{Planck}, is much larger than the known baryonic fraction
$\Omega_b = 0.049$.
Dark matter is present as a gravitational effect on all scales: galaxies, small and
large galaxy groups, clusters and superclusters, CMB anisotropies over the full horizon,
affecting baryonic density oscillations over large scales, and cosmic shear in the large-scale
matter distribution. The correct nature or explanation of dark matter is not
known, whether it implies unconventional particles or modifications to gravitational
theory, but gravitational effects have convincingly proved its existence in some form.
In the standard Cold Dark Matter model (CDM), dark matter appears to be collisionless:
it appears not to interact with baryonic matter except by gravitation. No other properties of DM are known.

Dwarf spheroidal galaxies (dSph) are the most DM-dominated objects known, they are even much cleaner from baryons than is the Galactic Center (GC). Among them one finds the smallest stellar systems known to contain dark matter, and they are usually exhibiting very high mass-to-light ratio values, $\Upsilon\equiv M/L$ from 10 to several 100. Here $M$ comprises the masses of baryonic and electroweak particles and dark matter taken together, and light $L$ luminous energy, radiated as photons by baryons and electroweak particles. The fainter an object is, the less luminous mass does it contain, and therefore the higher must its DM content be to prevent tidal disruption by the gravitational potential of its surrounding galaxy or cluster. Comparing the properties of dwarf galaxies in both the core and outskirts of the Perseus Cluster, Penny and Conselice \cite{Penny} found such a clear correlation between $\Upsilon$ and the luminosity of the dwarfs.

It is the purpose of this study to investigate whether the indirect study of DM halos in dSphs would reveal something about the nature of DM.

\section{The morphology of dwarf galaxies}

In the simplest dynamical framework one treats massive systems (galaxies, groups,
clusters) as statistically steady, spherical, self-gravitating structures of $N$ objects with
some average mass $m$ and average velocity $v$ or velocity dispersion $\sigma$. Gas cooling
and star formation within DM halos is now considered the standard paradigm for the
origin of galaxies. The baryonic content in galaxies is affected by these processes.
Simulations of different cosmological scenarios can make detailed predictions of the
internal structure of the invisible galactic halos.

Morphologically, dSphs (including a few dwarf ellipticals) can be divided into two main types:

- Blue compact dwarfs (BCD) with bright, compact, star-bursting regions, and

- Gas-rich dwarfs with multiple irregularly distributed H\i\i\ regions that are forming
stars at a relatively-low rate, named dwarf irregulars (dIrr). The latter are often very
poorly approximated by a spherical form, and therefore average quantities such as
mass, velocity, and velocity dispersion are more easily influenced by their environment
than for more massive galaxies.

An extreme type of systems are the ultra-faint dwarf galaxies (UFD). When
interpreted as steady state objects in virial equilibrium they would be the most DM
dominated objects known in the Universe. Their half-light radii range from 70 pc to
320 pc.

In contrast, ultra-compact dwarf galaxies (UCD) are stellar systems with masses
of around $10^7$ -- $10^8 M_{\odot}$ and half-mass radii of 10 -- 100 pc. A remarkable
properties of the UCDs is that their dynamical mass-to-light ratios, $\Upsilon$, are on average only about twice as large as those of globular clusters of comparable metallicity, and they also
tend to be larger than what one would expect based on simple stellar evolution models.
UCDs appear to contain very little or no dark matter.

An additional type within the class of low-surface brightness (LSB) galaxies are extremely metal-poor (XMP) blue diffuse dwarfs (BDD) embedded in a BCD continuum. James \& al. have reported \cite{James} a population of 12 such high-$z$ objects which appear
to be intermediate between the BCD and the typical actively star-forming dIrr
populations. They exhibit ongoing star-formation and surface brightness fluxes such
that they may be either active or quiescent. However, to conclude that 12 objects are excellent analogues for galaxies in the early Universe is statistically unconvincing, we don't have an early Universe to compare with.

On the scale of the Milky Way (MW) there are large statistical uncertainties encountered in the stochastic formation of MW-like halos, even within simple CDM models not including
baryonic processes .

Going beyond the MW one is tempted to draw on results from Local Group (LG) studies to explain the evolution of galaxies in the broader universe, although the
feeble statistics does not warrant that LG dwarf galaxies are representative of all
dwarf galaxies \cite{Weisz}.

In general, due to the small number of dwarf galaxies known and well measured
their morphological classification may not be very reliable or useful.

\section{The cusp-core problem}

Since DM and the shapes of DM halos cannot be seen, they need
to be simulated or fitted by empirical formulae. Because baryons dissipate energy and
so collapse to smaller scales than DM, they constitute a sizeable fraction of the mass in
the central regions of all but the faintest galaxies \cite{Bell}. The stellar component is uncertain in the mass-to-light ratio, as is the gas component in the conversion from neutral
hydrogen to total gaseous mass. Measured gas velocity fields have to be translated
into velocity curves, which is not possible for dIrrs.

Starting from a smooth early Universe the Cold DM model (CDM) has been quite
successful in reproducing the evolution into the cosmic structures observed today, it
prevails to all alternative models and it explains perfectly the observed fluctuations
of the CMB and the structures on large scales \cite{White}. Mostly the halo shape of dSphs is taken to be spherically symmetric so that the total gravitating mass profile $M(r)$ depends on only three conventional parameters: the mass proportion in stars, the halo mass and the length scale. DM is not required to have any unconventional properties like self-interaction or right-out new physics, it interacts only gravitationally.

The density profile of large spiral galaxies is mostly well described by the Navarro-Frenk-White \cite{NFW} (NFW) empirical formula with $\rho\propto 1/r$. The NFW-simulations thus predict a central cusp or bulge. In contrast to such simulations, however, many dwarf galaxies and low surface brightness (LSB) galaxies appear to have almost constant-density cores, with density profile $\rho\approx r^{-0.2}$. The rotation curves are often slowly rising with $r$ out to the outermost observable point where they do not yet constrain the maximum circular velocity. This disagreement is referred to as the \textit{cusp-core} problem.

However, the situation need not be considered that problematical, for several reasons reasons discussed below.

\subsection{Measurements}

Only relatively nearby dwarfs can be studied in any detail, and redshift is then a poor distance measure.

The MW satellites are not field galaxies. To infer the mass distribution of
their DM halos is significantly harder than for field galaxies when even the MW halo mass is not known to better than a factor of $\pm 2$.

Sculptor, a dSph of stellar
mass 107 $M_{\odot}$ and located $\approx 80$ kpc from the Galactic Center is a particularly interesting case. Modeling it in various ways, taking the available stellar data to be sampled
from a single stellar population and assuming spherical symmetry, Lovell \& al. \cite{Lovell} and Strigari \& al. \cite{Strigari} conclude that the kinematical data are consistent with an NFW halo, but also allow a core, so all interpretations remain contentious. These conflicting conclusions reflect differences in analysis methods rather than in observational data, but also the lack of statistics. The observed stellar distribution is clearly non-circular on the sky,
and Sculptor orbits within the potential of the Milky Way, so the effective potential
seen by its stars is time-varying.

\subsection{Statistics}

In general the number of dwarf galaxies which can be studied is very small and the populations do not lend themselves to proper statistical analysis. Three compilations offer examples of this.

Verbeke \& al. \cite{Verbeke} discuss 10 simulated faint, gas-dominated, isolated, star-forming dwarf galaxies, comparing them with three real objects, Leo P, Pisces A, and Leo T. There are tensions between simulations and observations which the authors claim to be effectively resolved when taking into account the regulating influence of the ultraviolet radiation of the first population of stars on a dwarf's star formation rate. The authors then confirm, for instance, that the simulations reproduce well the observed relation between stellar mass and neutral gas mass inside a $3\sigma$ prediction interval. My question is: what does $3\sigma$ mean for a population of 3 galaxies?

Oh \& al. have reported \cite{Oh} a study 7 field dwarfs from the THINGS compilation, finding $\rho\propto r^{-0.4\pm 0.15}$ in the innermost region (the baryonic contribution is here not subtracted). However, a sample of 7 does not form a Gaussian distribution, is too small to be quoted with $1\sigma$ statistical errors. A Poisson distribution with the variance equal to the mean would be more in place.

The 77 dwarf galaxies compiled by Oman \& al. \cite{Oman} were selected to have rotation
curves extending to at least twice their stellar half mass radius. Their distribution in the plane of the galaxy formation efficiency $f_{eff}$ versus $V_{circ}$ (figure 2) shows such an even spread that no obvious functional dependence can be statistically well-fitting. In figure 3 six selected dwarfs are being compared: half of them have rotation curves with signs of an inner core, whereas the other half are reasonably well fit by cuspy NFW profiles, thus the situation is inconclusive.

In general the NFW simulations significantly over-predict the central density of DM in dwarf galaxies \cite{Moore, de Blok}, so that the radial densities are better fitted by an Einasto profile \cite{Einasto} or a Burkert profile \cite{Burkert} .

\subsection{Cored profiles can be flattened cusps}

Hierarchical formation requires that dwarf galaxies should have been the oldest
systems forming stars. It is natural to assume that the central shape should in general have have been of the cusp- or bulge-form, as they come out in the NFW simulations. This form would subsequently, over a time scale of a few Gyr, has been flattened by one or several processes.

Given the scarcity of dwarfs, their faint
luminosity, and the long time perspective back to their formation, it is a hopeless
task to select which processes have contributed to reducing the central DM density.
The literature already knows a large choice of conventional processes which do not require
knowledge of the physics of dark matter particles, cf. eg. the review of A. Pontzen \&
F. Governato \cite{Pontzen}. The central cusp could have been flattened by:

-Protogalaxy merger events \cite{Olszewski, Coleman, Yozin}.

-Galactic centers decay towards the centers of their host galaxies, where they
collide and merge.

-Star formation activity depending on the stellar to halo mass ratio, even as
late as at $z\approx 9-11$ or $t\approx 10$ Gyr, could have driven gas out of dwarf galaxies.
Small halos do not form stars nor host galaxies due to baryon depletion and cosmic
reionization feedback. It is yet unclear even to what extent the available energy in
stellar populations couples to the gas through heating and radiation pressure. Inferring the dark matter density requires high spatial resolution of the gas and stellar kinematics.

-Tidal interactions and dynamical friction with a larger galaxy or a BH.

-Formation of nuclear clusters or cores that reduce the central density and depress
systematically local estimates of the circular velocity by altering the structure of the
DM halo and by depressing local estimates of the circular velocity \cite{Navarro, Mashchenko}.

-Energy liberated from star populations can be large compared to the binding
energy of the galaxies. Removing most of the baryons in a rapid, dramatic starburst
event could over-compensate for a previous adiabatic contraction. In particular
repeated outflow episodes interspersed by re-accretion may have a cumulative effect.

-Galactic winds constituted by in-flowing matter such as metal-enriched gas
moving at hundreds of kilometers per second in bubbles extending to 100 kpc or
more, but not attaining the escape velocity, thus accreting or re-accreting matter.

-Processes providing the energy to halt collapse are collectively named \textit{feedback}
\cite{Croton, Bower, Stinson, Hopkins}, and include the energetics of star formation, supernova winds, radiation from young stars, cosmic ultraviolet background, radiation and heat from black hole accretion, blow-out of gas. Spectroscopic observations reveal the ubiquity of massive galaxy outflows driven by various feedback processes \cite{Shapley, Weiner, Martin}.

-There is a lack of any
clear analytic framework for understanding the behaviour of gas and the apparently
irreversible response of DM to astrophysical processes over cosmic time.

If sufficient energy could be given \textit{gravitationally} to dark matter particles in the
center of the halo, they would then migrate outwards, reducing the central density. Where this energy should come from requires an unconventional mechanism, or new physics.

\section{Conclusions}

There is an ubiquity of conventional processes that could have flattened cusped dwarf galaxies. It is also possible that several processes have contributed in ways which makes it impossible to disentangle them. But there are no compelling reasons to introduce elements of new physics for explanation, such as self-interactions, several types of DM particles or dark energy.

It is clearly not avoidable that this field exhibits a lot of cosmic variance with potentially catastrophic consequences for the CDM model \cite{Governato}. Nevertheless, statistics is also a science with rules about what can be concluded.

\end{document}